\documentclass[amsmath,amssymb,twocolumn]{revtex4}

\def\prb{{\textit{ Phys. Rev. B,} }}

\usepackage{pslatex,revsymb,bm,latexsym,times,amssymb,amsmath,graphicx,epsfig}

\newcommand{\ud}{\mathrm{d}}

\newcommand{\etal}{\textit{et al.}}
\begin{document}
\title{Spin relaxation in isotopically purified silicon quantum dots}
\author{M. Prada$^{1,2}$, R. H. Blick$^{2}$, and R. Joynt$^1$},
%\author[pur]{G. Klimeck}
\affiliation{$^{1}$Department of Physics, 1150 University Ave., University of Wisconsin-Madison, Wisconsin 53706 (USA)}
\affiliation{$^{2}$Electrical and Computer Engineering, University of Wisconsin-Madison, Wisconsin 53706 (USA)}

\begin{abstract}
We investigate spin-flip processes of Si quantum dots due 
to spin-orbit coupling. % in the absence of a parallel magnetic field. , $B_\parallel$. \ 
We utilize the spin-orbit coupling constants related to bulk and structure 
inversion asymmetry obtained numerically for two dimensional heterostructures. \ 
We find that the spin-flip rate is very sensitive to these
coupling constants. \ 
%We use the  quantitative nanoelectronic modeling tool NEMO-3D that allows to 
%incorporate the effects of the interfaces to atomic scale. \ 
We investigate the nuclei-mediated spin-flip process and find the level 
of $^{29}$Si concentration for which this mechanism become dominant. \ 
\end{abstract}

%\begin{keyword}
%Spin-orbit \sep silicon \sep heterostructures \sep tight-binding \sep Rashba \sep Dresselhaus



% MSC codes here, in the form: \MSC code \sep code
% or \MSC[2008] code \sep code (2000 is the default)

%\end{keyword}

% main text
\maketitle
\section{Introduction}
The control of single-electron spin in nanoscale devices is a key element in the
field of spintronics, where the spin degree of freedom is used for information
transfer and processing. \
The vast majority of proposed spin devices are based on semiconductors 
\cite{kane,friesen,friesen2,datadas}, where single-electron spin has been probed to
be controllable by means of voltages applied to electrostatic gates \cite{petta}.
A promising technology for the implementation of quantum computation (QC)
involves the storage of quantum information in the spin of electrons in
quantum dots (QDs). \ The key requirement is that the lifetime of the spins
is long compared with the time required for the operation of logic gates. \
This has motivated the development of QDs in Si-based materials \cite{nakul}, where spin-orbit
coupling (SOC) is weak and isotopic enrichment can eliminate hyperfine coupling (HC). \
In such scenario, processes limiting QC are dominated by
SOC resulting from spatial inversion asymmetry. \ 
In typical solid state systems, macroscopic electric fields cause structure inversion
asymmetry (SIA), giving rise to Rashba-type terms \cite{rashba} of the form: 
\[
H_\mathrm{R}= \alpha (\hat \sigma_x \hat k_y -\hat \sigma_y \hat k_x),
\] 
whereas fields resulting from the lack of an inversion center lead to 
bulk inversion asymmetry (BIA) and to the Dresselhaus term,  
\cite{dresselhaus}: 
\[
H_\mathrm{D} = \beta (\hat \sigma_x \hat k_x -\hat \sigma_y \hat k_y).
\] 
Here, $\sigma_i$ and $ k_i$ denote spin and momentum operators, respectivelly, and 
$\alpha$, $\beta$ are the Rashba and Dresselhaus coupling constants, respectivelly. \
Although both SOC contributions have been noted for decades, 
their absolute value 
have been measured simultaneously only recently \cite{ensslin}. \
The different theoretical models show controversy \cite{hu},
with debated estimations for Rashba or Dresselhaus related parameters. 

In this paper we calculate spin-flip rates for typical Si QD via higher-energy 
virtual state, involving also phonon emission. \
We define the regime for which SOC is the dominant
source of relaxation as a function of $^{29}$Si isotope concentration.  \ 
We utilize in our calculations the SOC parameters extracted numerically 
using for first time a  model that includes atomic crystal
symmetry, spin and interfaces built into the basis representation \cite{nextprl}, 
that aims to clarify the existing controversy. 

%Understanding the processes that relax
%spins can point to strategies for minimizing relaxation and coherence times,
%thereby improving coherent control of quantum systems. \ \ 

This paper is organized as follows: 
Sec. \ref{sec1} contains a description of the model: 
we describe the electron-phonon interaction, and the spin-flip 
mechanism, namely SOC and hyperfine coupling (HC).  
In Sec. \ref{sec2} we present our results, 
and finally, Sec. \ref{sec3} is devoted to conclusions.  

\section{Methods}
\label{sec1}
\subsection{Model}
\begin{figure}[!hbt]
\centering\includegraphics[angle=0, width = .2\textwidth]{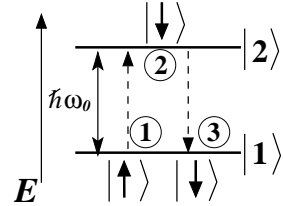}
\caption{\footnotesize 
Schematic representation of a spin-flip process for a Si QD with one electron. 
}
\label{figscheme}
\end{figure}

We consider a QD formed in a two dimensional (2D) heterostructure
with parabolic confinement. \ 
Spin-flip is considered via an
orbital (or valley) state with energy 
$\hbar\omega_0$, as depicted in Fig. \ref{figscheme}. \  

In order to calculate the relevant transition rates for spin relaxation, 
we consider a perturbation given in general by
$\delta H = H_{\mathrm {sf}} + H_{\mathrm{ph}}$,
$H_{\mathrm{ph}}$ corresponding to the electron-phonon
coupling, and  $H_{\mathrm {sf}}$ to the mechanism causing a spin
flip in the dot. \ The amplitude of a 
spin-flip event in the ground state then is given in second order
perturbation theory,
\begin{equation}
\label{eq:st}
\langle 1_\uparrow | \delta H |1_\downarrow \rangle \thickapprox 
\sum_n\left(
 \frac{\langle 1_\uparrow | H_{\mathrm{ph}} |n_\uparrow \rangle \langle n_\uparrow| H_{\mathrm{sf}} |1_\downarrow\rangle}
     {E_n  - E_1 + \Delta } %+ \hbar \omega_n)}
     %\right) 
     %+
%\sum_r\left(
 %\frac{\langle 1,\downarrow | H_{\mathrm{sf}} |n,\uparrow \rangle \langle n,\uparrow| H_{\mathrm{ph}} |1,\uparrow \rangle}
      %{E_{r} - E_1 -\Delta} %+ \hbar \omega_n)}
      \right),
\end{equation}
where  $\Delta$ is the energy exchanged with the bath.  \ Here, 
$\uparrow,\downarrow$ denote the spin state and $n$ is the orbital degree of freedom. \ 
The orbitals are described in terms of Fock-Darwin states. \  
The ground-state spin-flip rate  $|1_\uparrow\rangle$ to $|1_\downarrow\rangle$ is then given by Fermi's
golden rule:
\begin{equation}
\label{eq:st2}
        \Gamma_{\uparrow,\downarrow} = 
        \frac{2\pi}{\hbar}|\langle 1\uparrow | \delta H |1\downarrow\rangle|^2 \delta(E_i-E_f).
\end{equation}
In this notation, $|1\downarrow\rangle$ denotes the initial state of electron, nuclei
and phonons, $|1\downarrow\rangle \to |i\rangle$ =
$|1\downarrow\rangle \otimes |i_n\rangle \otimes |i_{\mathrm{ph}}\rangle$, likewise
$|1 \uparrow\rangle \to |f\rangle$ =
$|1 \uparrow\rangle\otimes|f_n\rangle \otimes |f_{\mathrm{ph}}\rangle$. \
Inserting Eq. (\ref{eq:st2}) into (\ref{eq:st}) we obtain (setting $\Delta\to$0):
\begin{equation}
\label{eq:st3}
\Gamma_{\uparrow,\downarrow} =
  \frac{2\pi}{\hbar}
   |\langle \langle f_{\mathrm{ph}} | H_{\mathrm{ph}} 
      |i_{\mathrm{ph}}\rangle|^2 \delta(E_i-E_f) 
   \frac{|\langle 1\uparrow|H_{\mathrm{sf}} |1\downarrow \rangle|^2}
    {(\hbar\omega_0)^2}.%\delta_{\downarrow^\prime}. %\nonumber \\
      %= \Gamma_{\mathrm{ph}}\times \zeta^2
      %\frac{|\langle s^\prime|H_{\mathrm{sf}} |T\rangle|^2}{(E_{T}-E_{s^\prime})^2} 
%\end{eqnarray}
\end{equation}
We note that the phonon and the spin-flip parts can be factorized, so we 
define $\Gamma_{\uparrow,\downarrow} = \Gamma_{\mathrm{ph}}\times \zeta^2$, with
\begin{equation}
\Gamma_{\mathrm{ph}} = \frac{2\pi}{\hbar}|\langle i | H_{\mathrm{ph}} |f \rangle|^2 \delta(E_i-E_f); \quad
\zeta^2 = \frac{|\langle 1\uparrow|H_{\mathrm{sf}} |1 \downarrow\rangle|^2}{(\hbar\omega_0)^2}%\delta_{\uparrow,\downarrow} .
\label{eq:gammaphonon1}
\end{equation}
Hence, $\Gamma_{\mathrm{ph}}$ and $\zeta$ can be treated
separately. \ $\Gamma_{\mathrm{ph}}$ describes the phonon-emission 
process, coupling two orbitals via electron-phonon deformation potential, 
and $\zeta$ includes the spin-flip process. We give next a detailed description 
of both terms. \\

\subsection{Phonon-electron coupling}
For Si under compressive stress along [001], the electron interacts
with a phonon of momentum $\bm{q}$ via deformation potentials \cite{charlie,
herring, hasegawa}. The Hamiltonian reads:
\begin{eqnarray}
\label{eq:Heph1}
H_{\mathrm{ph}} &=& \sum_s\sum_{\bm q} 
  i[a^*_{qs}e^{-i{\bm q r}} + a_{qs}e^{i{\bm q r}} ]
  q(\Xi_d \hat e_x^s\hat q_x + \Xi_d \hat e_y^s\hat q_y + \nonumber\\&& 
  (\Xi_d + \Xi_u)\hat e_z^s\hat q_z)
\end{eqnarray}
where $\langle n_q-1| a_q | n_q \rangle = \sqrt{(\hbar n_q/2M_c\omega_q)}$,
$M_c$ is the mass of the unit cell and $\hbar\omega_q$ the phonon energy. \
Here, $s$ denotes the polarization of the phonon (two transverse and one
longitudinal), $\bm q$ is the wavevector, $\Xi_u$ and $\Xi_d$ are the
electron-phonon coupling parameters in Si (see table \ref{table1}). \ This
is slightly simpler than the corresponding Hamiltonian in GaAs because of
the absence of the piezoelectric coupling in Si. \
% ($\Xi_u$ = 9.29 eV, $\Xi_d$ = -10.7eV). \
In order to calculate $\Gamma_{\mathrm{ph}}$, we use the
electric dipole (ED) approximation, $e^{i{\bm q r}} \thickapprox 1 + i {\bm q r}$ (valid in
the range of ~0.1-0.2 meV, which is roughly the phonon energy involved in
this relaxation process). \\ 
For the process depicted in Fig. \ref{figscheme}, mixing 
of $s$-orbitals occur at 0-th order in ED (to lowest order). \ We summarize in table
\ref{table1} the parameters relevant for the computation of $\Gamma_{\uparrow,\downarrow}$
used throughout the text.
\begin{table}[!hbt]
\centering
\begin{tabular}{|c|c|c|c|c|c|c|}
  \hline
   $\Xi_u$[eV]& $\Xi_d$[eV] & $\rho_{\mathrm {Si}}[$kg$/$m$^{3}$]& $v_l$[m/s] &   $v_t$[m/s]
   &$A$  [$\mu$eV$\cdot$nm$^{3}$] & $\langle\rho^2\rangle$[nm$^2$]
   %&$\epsilon_{\mathrm{ST}}$
\\
  \hline
   9.29 &-10.7  & 2330 &9330&5420 
   & 0.2&4$\times$10$^2$%& 0.2meV
\\
  \hline
\end{tabular}
\caption{\footnotesize Physical constants and material parameters for Si
and the QD as in \cite{nakul}.}
\label{table1}
\end{table}
Using the 0-th order ED approximation,
$e^{i{\bm q r}} \thickapprox 1 $, we obtain
 for the electron-phonon  coupling:
\[
\widehat H_{\mathrm{ph}} \thickapprox
\sum_q(a_q + a_q^\dag)|q| 
[\Xi_d\hat e_xq_x +\Xi_d\hat e_yq_y +(\Xi_d + \Xi_u)\hat e_zq_z], 
\]
which, using (\ref{eq:gammaphonon1}), gives the rate:
\begin{eqnarray}
\label{eq:matrixeldir}
\Gamma_{\mathrm{ph}}^{\mathrm{D}}
%|\langle i_{\mathrm{ph}}|\widehat H_{\mathrm{ph}}^{\mathrm{D}} 
%|f_{\mathrm{ph}} \rangle|^2 
&\thickapprox &
\frac{(n_q + 1)}{2\rho_{\mathrm {Si}}(2\pi)^2} 
\sum_s\int_0^{2\pi}\ud\varphi \int_0^\pi \ud(\cos{\theta})\int_0^\infty \ud q_s
\frac{q_s^4}{\omega_{q_s}} 
[\Xi_d\hat e_x^s\hat q_x + \nonumber\\ 
&& \Xi_d\hat e_y^s\hat q_y 
+(\Xi_d + \Xi_u)\hat e_z^s\hat q_z]^2 \delta(\hbar\omega_{q_s}-\hbar\omega_0).
\end{eqnarray}
Here,  $n_q$+1 $\thickapprox$ 1 in
the range of low temperatures considered, and $\rho_{Si}$ % = 2330Kg$\cdot$m$^{-3}$ 
is the density of Si (see table \ref{table1}). \
$\hat e_i$ is the $i$-th component of the unitary polarization vectors
and $s$ denotes the polarization, %(see (\ref{eq:polarvecs})),  
$s$=l, t$_1$, t$_2$:
\begin{eqnarray}
\label{eq:polarvecs}
\hat e^{l} & = & \sin{\theta} \cos{\varphi}\hat u_x +  
        \sin{\varphi} \hat u_y -\cos{\theta} \cos{\varphi}\hat u_z \nonumber \\
\hat e^{t_1} & = & \sin{\theta} \sin{\varphi}\hat u_x -
        \cos{\varphi} \hat u_y -\cos{\theta} \sin{\varphi}\hat u_z \nonumber \\
\hat e^{t_2} & = & \cos{\theta} \hat u_x + \sin{\theta} \hat u_z.
\end{eqnarray}
and $\hat q = (\sin{\theta} \cos{\varphi}, \sin{\theta} \sin{\varphi},\cos{\theta})$.
To integrate Eq. (\ref{eq:matrixeldir}), we assume an isotropic phonon
spectrum, $E_{\mathrm{ph}}= \hbar \omega_{qs}$ with a dispersion relation
$\omega_{qs} = v_s q$, $v_s$ being the sound velocity of the mode $s$. This
gives: 
\[
\int_0^\infty \frac{dq q_s^4}{\omega_{q_s}} 
        \delta(\hbar\omega_{q_s}-\hbar\omega_0) = 
%\frac{1}{\hbar^4 v_s^5} \int_0^\infty 
%\ud(\hbar\omega_{q_s})(\hbar\omega_{q_s})^3
        %\delta(\hbar\omega_{q_s}-\epsilon_{\mathrm{ST}}) = 
\frac{(\hbar\omega_0)^3}{\hbar^4 v_s^5}.
\]
Using the above result, the polarization vectors
of (\ref{eq:polarvecs}), rearranging (\ref{eq:matrixeldir}), and performing
the angular integral, we obtain a compact expression for the rate:
\begin{eqnarray}
\label{eq:gammaphD}
\Gamma_{\mathrm{ph}}^{\mathrm{D}} 
%=\frac{2\pi}{\hbar}
%|\langle i_{\mathrm{ph}}|\widehat H_{\mathrm{ph}}^{\mathrm{D}}|f_{\mathrm{ph}} 
%\rangle|^2\delta(\hbar\omega-\epsilon_{\mathrm{ST}})
&\thickapprox&
\left(\frac{\hbar\omega_0}{\hbar}\right)^3 \frac{1}{2\pi\hbar \rho_{Si}}\frac{1}{15}
\left[  
\frac{15\Xi_d^2 + 10\Xi_d\Xi_u + 3\Xi_u^2}{v_l^5} +\right.\nonumber\\ &&\left.
 \frac{3\Xi_d^2 + 4\Xi_d\Xi_u + 2\Xi_u^2}{v_t^5}
\right].
\end{eqnarray}

\subsection{Spin orbit coupling}
As pointed out before, spin-flips can be provided by two different mechanisms,
the HC with the $^{29}$Si nuclei and the SO coupling. \
For an isotopically-purified sample, we expect SO to be the dominant 
spin-flip mechanism.  \ Hence, we first evaluate SO mediated spin-flip 
and then consider the $^{29}$Si concentration at which the HC mediated 
spin-flip rate becomes comparable to the SO mediated one.  \  \

%A magnetic field can be applied perpendicular to the 2D system. \ 
We express the Hamiltonian in the convenient phase coordinates ($q_1$,$q_2$,$p_1$,$p_2$)
\cite{galkin}:
\begin{equation}
\label{eq:hosc2}
\hat H_{0} = \frac{1}{2m}
\left(
\hat p_1 ^2 + \hat p_2^2\right) + 
\frac{m^*}{2}(\omega_1^2\hat q_1^2 + \omega_2^2\hat q_2^2) +
\frac{1}{2}g\mu_BB\hat\sigma_z, 
%\frac{1}{2}m(\omega_{x}\hat x^2 + \omega_{y}\hat y^2)  \sigma_zg\mu_BB,%+\left(\frac{eB}{2c}\right)^2
\end{equation}
with $\omega_{1,2}=(\omega_0^2+\omega_c^2)^{1/2} \pm \omega_c$ and $\omega_c = eB/m $. \  
%where $u(\vec r)$ is the confining potential. \ 
\begin{figure}[!hbt]
\centering\includegraphics[angle=0, width = .35\textwidth]{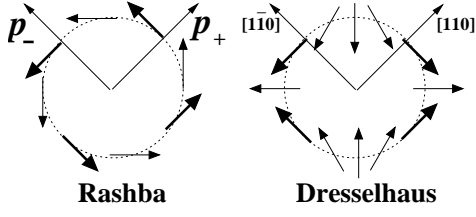}
\caption{\footnotesize 
(a) Schematic representation of a spin-flip process for a Si QD with one electron. 
(b) Spin direction for the eigenfunctions of Rashba and Dresselhaus contributions:
for $\bm{p}\parallel[110], [1\bar10]$, the effective magnetic fields are parallel.
}
\label{figrot}
\end{figure}

We add as a perturbation the Rashba and
the Dresselhaus SOC terms in a convenient basis rotated by 45$^\circ$ 
(see Fig. \ref{figrot}):
\begin{equation}
\label{eq:soc_r_d}
\hat H_{\mathrm{SO}} =\frac{1}{\hbar}\left[ 
(\beta + \alpha)\hat\sigma_+ \hat p_- +
(\beta -\alpha) \hat \sigma_-\hat  p_+
\right].
\end{equation}
We note that the SOC term does not commute with $\widehat H_0$. \
This implies that the spin does not remain fixed during the motion of the electron,
apart for an apparent homogeneous shift of the momentum to first order, which would
not change observables (other than just a shift in the energy levels). \ 
To get rid of such terms fixed by gauge invariance,
it is convenient to apply the transformation of Aleiner {\it{et al.}}, \cite{aleiner}
$\widehat U = \mathrm{exp}\{i\gamma_1\hat q_+\hat \sigma_- + i \gamma_2\hat q_-\hat \sigma_+ \}$,
($\hat q_\pm$ is the position operator along the directions $\hat e_\pm$), 
which allows to see how the levels split in the presence of $\widehat H_{SO}$. \ 
Using the  Zassenhaus formula, 
$e^{\hat A + \hat B} \simeq e^{\hat A } e^{\hat B} e^{-\frac{1}{2}[\hat A,\hat B]}\cdots$
we factorize $U$ to second order in SOC:
\begin{equation}
\label{eq:U1}
\widehat U = e^{i(\gamma_1\hat q_+\hat \sigma_- +  \gamma_2\hat q_-\hat \sigma_+ ) } \simeq
e^{i\gamma_1\hat q_+\hat \sigma_-   } e^{i \gamma_2\hat q_-\hat \sigma_+ }
e^{-\frac{1}{2}\gamma_1\gamma_2\hat q_+ \hat q_-[\hat\sigma_-,\hat \sigma_+]  }. %\dots =
%e^{i\gamma_1\hat x\hat \sigma_y   } e^{i \gamma_2\hat y\hat \sigma_x }
%e^{i\gamma_1\gamma_2\hat x \hat y\hat \sigma_z  }\dots,
\end{equation}
Retaining terms to third order in SOC, the momentum transform as:
\[
\widehat U^\dag \widehat P_+ \widehat U \simeq
\widehat P_+ + \hbar\gamma_1\hat \sigma_- -\hbar\gamma_1\gamma_2\hat q_- \hat\sigma_z -
2\hbar\gamma_1^2\gamma_2\hat q_-\hat q_+ \hat\sigma_+,
\]
and likewise: 
\[
\widehat U^\dag \widehat P_- \widehat U \simeq
\widehat P_- + \hbar\gamma_2\hat \sigma_+ +\hbar\gamma_1\gamma_2\hat q_+ \hat\sigma_z -
2\hbar\gamma_1\gamma_2^2\hat q_-\hat q_+ \hat\sigma_-.
\]
so that the total Hamiltonian, $\widehat H = \widehat H_0 + \widehat H_{\mathrm{SO}}$, transforms as
$\widehat H^\prime = 
\widehat U^\dag \widehat H \widehat U$. \ Considering terms up to third-order in SOC:
\begin{eqnarray}
\label{eq:htransformed2}
\widehat H^\prime &\simeq&
%\frac{1}{2m}( \widehat P_{x}^2 + \widehat P_{y}^2 ) + u(\vec r) +
\widehat H_0+
\frac{\hbar}{m l_1 l_2}[\hat q_+ \hat p_- - \hat q_- \hat p_+ ]\hat \sigma_z +
\nonumber\\&&
\frac{2\hbar}{m l_1 l_2}\hat q_+ \hat q_- 
\left[
  \frac{\hat p_+\hat \sigma_+}{l_1} - \frac{\hat p_-\hat \sigma_-}{l_2}
\right]= \widehat H_0+\widehat H_1+\widehat H_2,
\end{eqnarray}
where we have defined $\gamma_1$=${m(\alpha -\beta)}/{\hbar} \equiv l_1^{-1}$ and 
$\gamma_2$ = ${-m(\alpha +\beta)}/{\hbar} \equiv -l_2^{-1},$ with 
$l_{1,2}$ characterizing the length scale associated with the strength of 
the SOC for electrons moving along the crystallographic directions. \ 
This choice allows us to cancel out the linear terms in momentum. \ 
The second term of Eq. (\ref{eq:htransformed2}) gives an effective magnetic 
field that can be
expressed as:  
\begin{equation}
\label{eq:h1}
\widehat H_1 = i h_1%[h_{1\bar{1}0} k_{1\bar{1}0} - h_{1{1}0}k_{1{1}0}]
        [\hat a_1\hat a_y^\dag - \hat a_y \hat a_x^\dag]\hat\sigma_z, \quad
\end{equation}
where we have defined: 
\[
h_1 = 2m^*(\alpha^2-\beta^2)\left(\sqrt{\frac{\omega_2}{\omega_1}}   
  +\sqrt{\frac{\omega_1}{\omega_2}}\right) .
\]
We note that $\widehat H_1 $ does not break the Kramers degeneracy, 
and hence, does not contribute to spin-flip. \ However, the last term of 
Eq. (\ref{eq:htransformed2}) does break Kramers degeneracy, allowing spin-flip 
terms. \ In second quantization, we have: 
%\[
%\widehat H_2 =
%\frac{2\hbar}{l_1l_2m}\hat x \hat y \left[\frac{\widehat P_y\sigma_y}{l_2} -\frac{\widehat P_x\sigma_x}{l_1} \right] =
%\frac{2i\hbar^2 }{m^2l_1l_2}\sqrt\frac{\hbar m}{2\omega_1 \omega_2 } [\hat a_x^\dag a_y + a_y^\dag a_x]
%\left[ \frac{\sqrt{\omega_2}}{l_2} (\hat a_y^\dag -\hat a_y)\hat\sigma_y -
       %\frac{\sqrt{\omega_1}}{l_1} (\hat a_x^\dag -\hat a_x)\hat\sigma_x \right].
%\]
\begin{eqnarray}
\label{eq:h22ndq}
\widehat H_2 &=& 
i\gamma_y[\hat a_x^\dag\hat a_x^\dag\hat a_y + %- \hat a_x\hat a_x\hat a_y^\dag +
\hat a_y^\dag (1-\hat n_x) - \hat a_y\hat n_x - \hat a_y^\dag  \hat a_x\hat a_x ]\hat \sigma_x \nonumber\\
&&-i\gamma_y [\hat a_y^\dag\hat a_y^\dag\hat a_x + %-\hat a_y\hat a_y\hat a_x^\dag +
\hat a_x^\dag (1-\hat n_y) - \hat a_x\hat n_y -\hat a_x^\dag\hat a_y\hat a_y ]\hat \sigma_y, 
%\quad \gamma_2 = m\sqrt{\frac{m}{\hbar\omega}}(\alpha^2-\beta^2).
\end{eqnarray}
where we have defined: 
\[
\gamma_y=(\beta-\alpha)^2(\beta+\alpha)m\sqrt{\frac{2m}{\hbar\omega_1}}; 
\ 
\gamma_x=(\beta+\alpha)^2(\beta-\alpha)m\sqrt{\frac{2m}{\hbar\omega_2}}.
\]
We then conclude that, for SOC, $\zeta_{\mathrm{SO}} \thickapprox \gamma_i^2/\hbar\omega$, 
from which we can proceed to calculate $\Gamma_{\uparrow,\downarrow}$, 
using (\ref{eq:h22ndq}) and (\ref{eq:gammaphD}) in (\ref{eq:gammaphonon1}). \ 
We emphasize that this term is third order in SOC, and applying Fermi 
Golden's rule, the spin-flip rate will appear in sixth order. \ Hence, 
$\Gamma_{\uparrow,\downarrow}$ is very sensitive to the SOC parameters, 
so it is critical to determine them to a high degree of 
precision. 

\subsection{Hyperfine Coupling.} 
For a electronic spin in a QD and in the presence of nuclear spins, a
contact interaction Hamiltonian can be assumed as a perturbation,
\begin{equation}
\label{eq:Hhf}
\widehat H_{\mathrm{hc}} = \sum_{i,j} \frac{4\mu_0}{3I}\mu_B\mu_I\eta
                {\bm S}_i {\bm I}_j\delta(\bm r_i - \bm R_j)=
                A\sum_{i,j}{\bm S}_i {\bm I}_j\delta(\bm r_i - \bm R_j)
\end{equation}
where ${\bm S}_i$ (${\bm I}_j$) and $\bm r_i$ ($\bm R_j$) denote the spin
and position of the $i^\mathrm{th}$ electron ($j^\mathrm{th}$ nuclei), and $A$ is the hyperfine
coupling constant. \ We now use (\ref{eq:Hhf}) as the spin-flip mechanism 
to calculate $\zeta$ of Eq. (\ref{eq:gammaphonon1}): 
\[
|\langle 1_\uparrow|H_{\mathrm{hc}} |2_\downarrow\rangle|^2 = 
A^2\sum_j|\langle \uparrow|S^+I_j^-\delta(r-R_j)|\downarrow\rangle|^2.
\]
Next we substitute the sum over $j$ by an integral over the nuclei with 
a density $C_n$: 
\[
|\langle 1_\uparrow|H_{\mathrm{hc}} |2_\downarrow\rangle|^2 =
A^2C_n\int\ud^3 R_j|\Phi_1(R_j)|^2 |\Phi_2(R_j)|^2 
\] 
which can be easily performed assuming Fock-Darwin states for $\Phi_i$, 
resulting:
\begin{equation}
\label{eq:result1}
\zeta_{\mathrm{hc}}=\frac{3A^2C_n}{8V_{\mathrm{QD}}(\hbar\omega)^2 }
\end{equation}
Using (\ref{eq:result1}) together with Eq.
(\ref{eq:gammaphD}) in  (\ref{eq:gammaphonon1}), we can thus obtain 
the spin-flip rate due to HC. %\begin{figure*}[!t]
Typical values for $\eta^{Si}$ of 186 have been reported \cite{shulman,abalmassov,paget},
yielding $A\thickapprox 2\times 10^{-7}$eV$\cdot $nm$^{3}$.
%$C_n$  is the effective nuclear density. 
Only about 4$\%$ of the nuclei have spin, so $C_n \thickapprox 
0.04\times 8/v_0 \thickapprox 2$nm$^{-3}$ ($v_0 \thickapprox 0.17$nm$^{3}$, is 
the unit cell volume for Si).
It is important to note that $\Gamma _{\uparrow\downarrow}$ is proportional to $%
C_{n}, $ i.e., to the total number of nuclei $N_{n}$ with which the
electrons interact. \ This is consistent with the simple picture that the
relaxation rate is proportional to the mean-square fluctuations in the
random hyperfine field. \ Formulas for spin relaxation rates due to
HC that give an apparent proportionality to $N_{n}^{-1/2}$
are common in the literature, and have given rise to the incorrect notion
that some sort of motional narrowing is at work. \ This is not possible,
since the fluctuations in the nuclear spin system are slow. \ In any case
the rate must vanish as $N_{n}\rightarrow 0.$ \ These formulas are correct,
but they generally involve other parameters that actually vary with $N_{n}.$ \ \

\section{Results}
\label{sec2}

We considered a typical QD formed as in \cite{nakul}
with a level separation of $\hbar\omega_0$ = 0.2meV. 
The $\alpha$ and $\beta$ parameters where extracted numerically using NEMO-3D 
\cite{nemo} on nanoHUB.org computational resources \cite{nanohub}.
In NEMO-3D, atoms are represented explicitly in the {\it{sp$^3$d$^5$s$^*$} }
tight-binding model, and the valence force field (VFF) method is employed to
minimize strain \cite{strain}. NEMO-3D enables the calculation of localized
states on a QW and their in-plane dispersion relation with a very high degree
of precision, allowing to extract the splittings along the in-plane directions
in momentum space \cite{nextprl}. \ 
For now, we note that the $\alpha$ value is well defined and depends on 
external electric fields applied to the sample, whereas the Dresselhaus, $\beta$, 
depends strongly on the atomistic details of the interface. \  

\begin{figure}[!hbt]
\centering\includegraphics[angle=0, width = .475\textwidth]{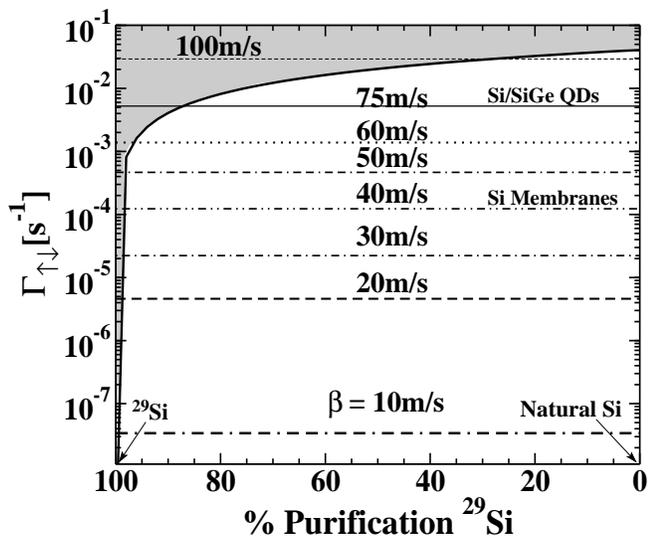}
\caption{\footnotesize 
 $\Gamma_{\uparrow\downarrow}$ as a function of  $^{29}$Si isotope
purification degree. \ The horizontal lines correspond to  $\Gamma_{\uparrow\downarrow}$
due to SOC, for various values of Dresselhaus coupling parameter, $\beta$. \ 
%The horizontal solid line shows $\Gamma_{\uparrow\downarrow}^{\mathrm{HC}}$ for 
}
\label{results1}
\end{figure}
Fig. \ref{results1} shows the calculated values for $\Gamma_{\uparrow\downarrow}$
via both mechanism, HC and SOC. 
The parallel lines correspond to $\Gamma_{\uparrow\downarrow}^{\mathrm{SOC}}$ for 
different possible values of $\beta$, 
whereas the thick solid line represents the spin-flip due to HC mechanism, 
$\Gamma_{\uparrow\downarrow}^{\mathrm{HC}}$, as a function of $^{29}$Si isotope
purification degree. \  
Using our numerical value for $\alpha$ and a few different values for $\beta$,  
we find that for natural Si, the HC dominates. 
We note that these depend strongly on the sample, so a Si sample with 
dominating SOC is possible. \   
In particular, for a Si/SiGe heterostructure, the $\Gamma_{\uparrow\downarrow}^{\mathrm{HC}}$
value corresponds to the horizontal solid line of Fig. \ref{results1}. We can 
see that a purification of about 80\% of the $^{29}$Si isotope will cause SOC 
to become dominant. \ In contrast, for a similar QD formed in a Si membrane, 
with purification of $\simeq$ 99\% we obtain SOC as the dominant mechanism, 
increasing the relaxation time by almost two orders of magnitude.

\begin{figure}[!hbt]
\centering\includegraphics[angle=0, width = .44\textwidth]{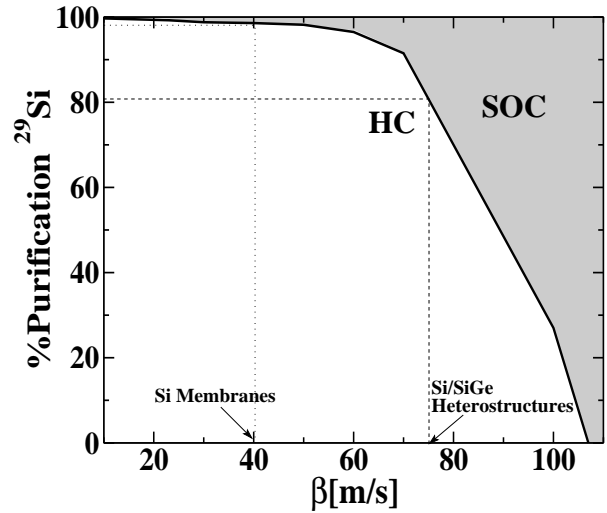}
\caption{\footnotesize Regions where SOC (shadded) or HC (light) are 
the dominant mechanism of relaxation as a function of Dresselhaus parameter, 
$\beta$, and degree of $^{29}$Si isotope purification: 
for small $\beta$, a larger isotopic purification degree is needed, and hence, 
larger relaxation times can be achieved. 
}
\label{results2}
\end{figure}

Fig. \ref{results2} represents the regions for which SOC or HC are 
the dominant spin-flip mechanism. \ It also summarizes the most relevant $\beta$
values found numerically: a typical QD defined in a Si/SiGe heterostructure 
has $\beta\simeq$75m/s, for which HC is dominant up to a purification level of 
81.5\%. A similar structure obtained on a Si membrane will show a larger Dresselhaus, 
and hence, larger purification will be needed to have SOC as dominating mechanism.

\section{Conclusions}
\label{sec3}

In summary, we have calculated the dominant rates for phonon-assisted
spin-flip relaxation for the ground state of a single-electron
 Si QD. \ $\Gamma_{\uparrow,\downarrow}$ is
found to be of the order of tens of seconds, very sensitive to SOC. 
We observe that choosing a sample with small SOC, as pure Si membranes, can lead to 
spin lifetimes of up to a few hours. \ 
QDs fabricated in Si/SiGe heterostructures are more limited by SOC, 
and purifications of only about 80\% would give spin lifetimes of a few minutes. 
\ Due to weak spin-orbit and HC, Si offers very long
coherence times, which are required for solid state qubits.

This work was supported by the Spanish
Ministry of Education and Science (MEC). We acknowledge support from
the Army Research Office (W911NF-04-1-0389) and the National Science Foundation (NSF-ECS-0524253) 
and through the MRSEC-IRG1 at UW-Madison. 

% \appendix

% \section{}
% \label{}
%\bibliographystyle{elsart-num}

\end{document}